\documentclass[traditabstract]{aa}
\usepackage[utf8]{inputenc}
\usepackage{amsmath}
\usepackage{amssymb}
\usepackage{microtype}
\usepackage{natbib}
\usepackage{graphicx}
\usepackage{txfonts}
\usepackage{color}
\usepackage{multirow}
\usepackage{xspace}
\usepackage{ulem}
\bibpunct{(}{)}{;}{a}{}{,} 
\newcommand{\ero}{{eROSITA}\xspace}
\newcommand{\art}{{ART-XC}\xspace}
\newcommand{\srg}{{SRG}\xspace}
\newcommand{\srca}{\object{SRGA\,J124404.1$-$632232}\xspace}
\newcommand{\srce}{\object{SRGU\,J124403.8$-$632231}\xspace}

\begin{document}

\title{\srca/\srce: a new X-ray pulsar discovered in the all-sky survey by SRG}
\author{V.\,Doroshenko\inst{1}, R.\,Staubert\inst{1}, C.\,Maitra\inst{2}, A.\,Rau\inst{2}, F.\,Haberl\inst{2}, A.\,Santangelo\inst{1}, A.\,Schwope\inst{3}, J.\,Wilms\inst{4}, D.A.H.\,Buckley\inst{5,6}, A.\,Semena\inst{7}, I.\,Mereminskiy\inst{7}, A.\,Lutovinov\inst{7}, M.\,Gromadzki\inst{8},
{L.J.\,Townsend \inst{5,9},
I.M.\,Monageng \inst{5,6}}}
\titlerunning{The new X-ray pulsar \srca/\srce}
\authorrunning{V. Doroshenko et al.}
\institute{Institut f\"ur Astronomie und Astrophysik, Sand 1, 72076 T\"ubingen, Germany
\and
Max-Planck-Institut f\"ur extraterrestrische Physik, Gie{\ss}enbachstra{\ss}e 1, 85748 Garching, Germany
\and
Leibniz-Institut f\"ur Astrophysik Potsdam (AIP), An der Sternwarte 16, 14482 Potsdam, Germany
\and
Dr.~Karl Remeis-Sternwarte and Erlangen Centre for Astroparticle Physics, Friedrich-Alexander-Universit\"at Erlangen-N\"urnberg, Sternwartstr.~7, 96049 Bamberg, Germany
\and
South African Astronomical Observatory, 
PO Box 9, Observatory Rd, Observatory 7935, South Africa
\and
Department of Astronomy, University of Cape Town, Private Bag X3, Rondebosch 7701, South Africa
\and
Space Research Institute (IKI) of Russian Academy of Sciences,
Prosoyuznaya ul 84/32, 117997 Moscow, Russian Federation
\and 
Astronomical Observatory, University of Warsaw, Al. Ujazdowskie 4, 00-478 Warsaw, Poland
\and
Southern African Large Telescope, PO Box 9, Observatory Rd, Observatory 7935, South Africa}


\abstract{Ongoing all-sky surveys by the the \ero and the {\it Mikhail Pavlinsky} \art telescopes on-board the Spectrum Roentgen Gamma (SRG) mission have already revealed over a million of X-ray sources. One of them, \srca/\srce, was detected as a new source in the third (of the planned eight) consecutive X-ray surveys by \art. Based on the properties of the identified optical counterpart it was classified as a candidate X-ray binary (XRB). We report on the follow-up observations of this source with Nuclear Spectroscopic Telescope Array (\textit{NuSTAR}), Neil Gehrels Swift Observatory (\textit{Swift}), and the Southern African Large Telescope (SALT), which allowed us to unambiguously confirm the initial identification and establish \srce as a new X-ray pulsar with a spin period of $\sim$538\,s and a Be-star companion, making it one of the first Galactic X-ray pulsars discovered by SRG.}
\maketitle

\section{Introduction}
The Spectrum Roentgen Gamma ({\it SRG}) mission was launched on July~13, 2019 from Baikonur, and started the all-sky survey on Dec~12, 2019. In total, eight complete scans of the celestial sphere, each lasting six months are foreseen to be conducted with each of the two instruments on board \srg, i.e. \ero\ \citep{2012arXiv1209.3114M,2021A&A...647A...1P} operating in the 0.2-10\,keV, and the {\it Mikhail Pavlinsky} \art \citep{2021arXiv210312479P} operating in the 4-30\,keV energy range.
Both instruments produce standard X-ray event data and produce not only images, but also spectra and light-curves for each point on the sky. At the end of the survey, the \ero\ surveys are expected to be a factor {20-40 more sensitive than the {\it ROSAT} All-Sky Survey in common energy band ($\sim2.5\times10^{-15}$\,erg\,cm$^{-2}$\,s$^{-1}$ for \ero \citep{2021A&A...647A...1P} compared to $\sim10^{-13}$\,erg\,cm$^{-2}$\,s$^{-1}$ \citep{2016A&A...588A.103B} for \textit{ROSAT}). Moreover, SRG will also conduct first ever true imaging all-sky surveys above the {\it ROSAT} pass band.}

A large majority of the several millions of new X-ray sources detected in the survey are expected to be extra-galactic objects (i.e. active galactic nuclei and galaxy clusters), and ordinary stars \citep{2012arXiv1209.3114M,2021A&A...647A...1P}. However, a significant expansion of the sample of the Galactic X-ray binaries can also be anticipated and will help to obtain important constraints on their luminosity function \citep{2013MNRAS.431..327L,2014A&A...567A...7D}. Of particular interest here are properties of the faint Be X-ray binaries (BeXRBs) which up to now have been mostly discovered with all-sky monitors during their strong outbursts. It is quite clear, however, that the population of known BeXRBs is potentially only the tip of an  iceberg as objects exhibiting fainter outbursts or no outbursts at all were likely missed due to the low sensitivity of all-sky monitors \citep{2014A&A...567A...7D}. In comparison with available wide-field instruments both \ero and \art are significantly more sensitive and are expected to unveil at least a part of this hidden population and thus facilitate advances in understanding of origin and evolution of the Galactic X-ray binaries.

On 2021 Jan 27, during the third consecutive all-sky survey, the {\it Mikhail Pavlinsky} \art telescope on board the \srg observatory detected a bright X-ray source \srca with a flux of ${\sim}10^{-11}\,\mathrm{erg}\,\mathrm{cm}^{-2}\,\mathrm{s}^{-1}$ in the 4--12 keV energy band \citep{2021ATel14357....1B}. 
Inspection of the \art data from previous two surveys in 2020 (see i.e. Table~\ref{tab:obs} for summary of the observations) revealed no excess at this position, so the source appeared to be new. However, it was found that an object was actually detected by \ero in all three surveys, which allowed to refine its X-ray position and identify a tentative optical counterpart of the X-ray source \citep{2021ATel14357....1B}. Near infra-red and X-ray colours, the X-ray flux, and an estimated X-ray luminosity of ${\sim}10^{34}$\,erg\,s$^{-1}$ implied by the {\it Gaia} distance of ${\sim}$5.8\,kpc \citep{2021AJ....161..147B}, strongly suggested that the object may be a new high-mass X-ray binary (HMXB). The variability of the counterpart, in particular a weak periodic, $\sim$138\,d, optical flux variations revealed by the Optical Gravitational Lensing Experiment \citep[OGLE,][]{2021ATel14361....1B} commonly observed also in other high-mass X-ray binaries with Be-star counterparts, strongly supported this suggestion. To confirm this identification, we observed the counterpart with the 
{Southern African Large Telescope  \cite[SALT;][]{2006SPIE.6267E..0ZB}}, revealing a prominent H$\alpha$ line in emission \citep{2021ATel14364....1B}, which is typical for Be X-ray binaries and also strongly supports the BeXRB nature of the source which motivated us to conduct follow-up observations of this X-ray source with \textit{NuSTAR}. 

In this paper we report in detail on the analysis of the observational data of \textit{NuSTAR}, SALT, and the \srg survey. The paper is organized as follows. In Sect. 2 we first briefly discuss observations used, and then discuss in detail analysis steps which led to identification of the near-infrared counterpart of X-ray source, and details of analysis for the follow-up \textit{NuSTAR} and SALT observations. We then discuss our findings in context of properties of other BeXRBs in Sect. 3, and finally summarize the results in Sect. 4.

\section{Observations and data analysis}
As already mentioned, there are two telescopes on board \srg, i.e. \ero and \art described in detail in \cite{2021A&A...647A...1P} and \cite{2021arXiv210312479P}, respectively. 
Due to the comparatively low exposures achieved up to now in the survey, data from both instruments were mostly used to obtain a precise position of the X-ray source, and measure flux variations between the three consequent surveys, whereas \textit{NuSTAR} data were used for detailed spectral and timing analysis.
Considering the non-detection of the source by \art in the first two surveys, we only use data from the third survey for this instrument. For \ero, data from all three surveys are used. 

\ero data were processed using the current (\texttt{c946}) eSASS (Brunner et al., submitted) pipeline configuration (used to characterize both software and calibration files version), which was also used to derive the source position and to extract X-ray spectra which were modeled jointly with \textit{NuSTAR} data to improve the low-energy coverage and constraints on the absorption column.
Details on calibration and in-orbit background of \ero are reported in \cite{2020SPIE11444E..4QD} and \cite{2020SPIE11444E..1OF} respectively. The data from the three all-sky surveys were analyzed independently, although spectrum of the source was assumed to be the same in all of them as described below. 

\art data were processed with the analysis software \texttt{ARTPRODUCTS} v0.9 with the calibration files version 20200401. A description of the \art telescope and the analysis software can be found in \citet{2021arXiv210312479P}. Taking into account the faintness of the source and the significantly shorter effective exposure for \art due to its smaller field of view, it was only possible to detect the source and estimate the number of counts above the background. Thus we use \art data mostly to improve constraints on the source flux by \ero in the third survey. In practice, we attributed all counts detected by \art to two broad energy bins covering 4-12 keV band, and modeled the \art data using the current response. Here we note also that even if \art and \ero observed the source at about the same time, difference in field of view of the two instruments and scanning pattern of the all-sky survey implies that the observations are not strictly simultaneous, so source flux measured by \art was estimated independently.

In addition to SRG data, a deep (55\,ks, obsid. 80660301002) observation of the source with \textit{NuSTAR} \citep{2013ApJ...770..103H} was obtained to characterize the broadband spectrum of the source and search for possible pulsations. The data reduction for this observation was done using HEASOFT v6.28 and caldb version 20210210 as described below. Soon after that, the Neil Gehrels \textit{Swift} Observatory also observed the object with XRT telescope (obsid. 00041978002), so we also use that observation. 

Furthermore, as part of a transient identification and characterisation program being undertaken at SALT \citep{Buckley2015},
we observed the optical counterpart on 2021 January 29 using the Robert Stobie Spectrograph (RSS) in low resolution mode ($R = 420$). In particular, two spectra, each with 1\,ks exposure covering the region 3300--9000\AA\xspace were obtained starting 23:34:05 UTC.
The summary of all observations is presented in Table.~\ref{tab:obs}.

Finally, to study properties of the optical counterpart, we used the archival OGLE $I$-band light curve published by \cite{2021ATel14361....1B} available via the OGLE XROM service\footnote{\url{http://ogle.astrouw.edu.pl/ogle4/xrom/srga_j124404.1-632232.html}}.  
\begin{table}[t]
    \centering
    \caption{Summary of observations of the source considered in this work.}
    \begin{tabular}{llll}
    \hline
         Instrument & T$_{\rm start}$, MJD & T$_{\rm stop}$, MJD &  Exposure, ks\\
         \hline
\ero & 58883.36 & 58885.86 & 0.501 \\ 
		& 59068.27 & 59070.43 & 0.380 \\ 
		& 59239.92 & 59241.59 & 0.289 \\
\art  & 59240.59 & 59241.42 & 0.172 \\
NuSTAR  & 59244.48 & 59245.53 & 54 \\ 
Swift/XRT & 59248.17 & 59248.77 & 1.7 \\ 
SALT & 59243.98 & 59244.01 & 2 \\
\hline
    \end{tabular}
    \label{tab:obs}
\end{table}

\subsection{X-ray position and near-infrared counterpart}
\begin{figure}
    \centering
    \includegraphics[width=0.9\columnwidth]{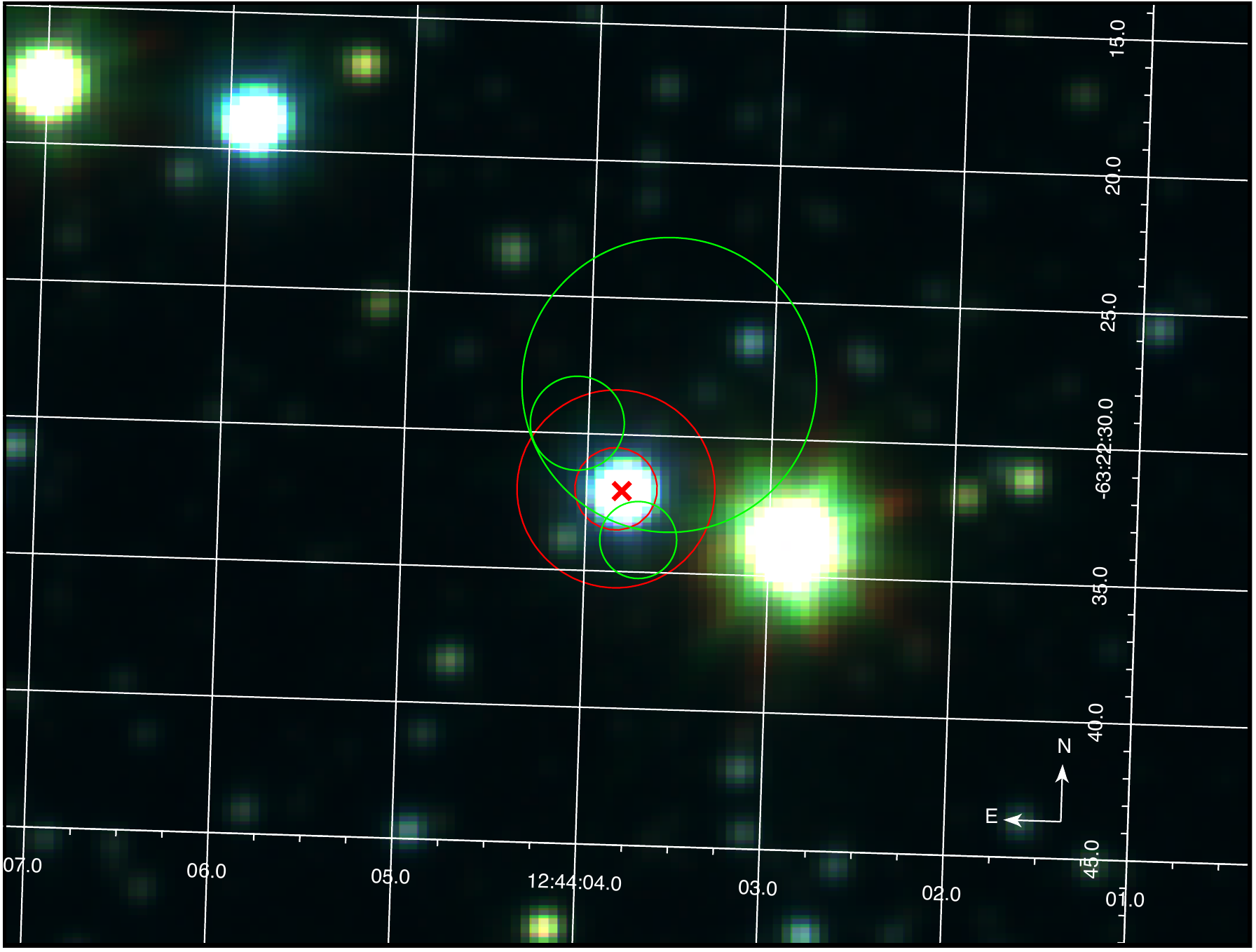}
    \includegraphics[width=\columnwidth]{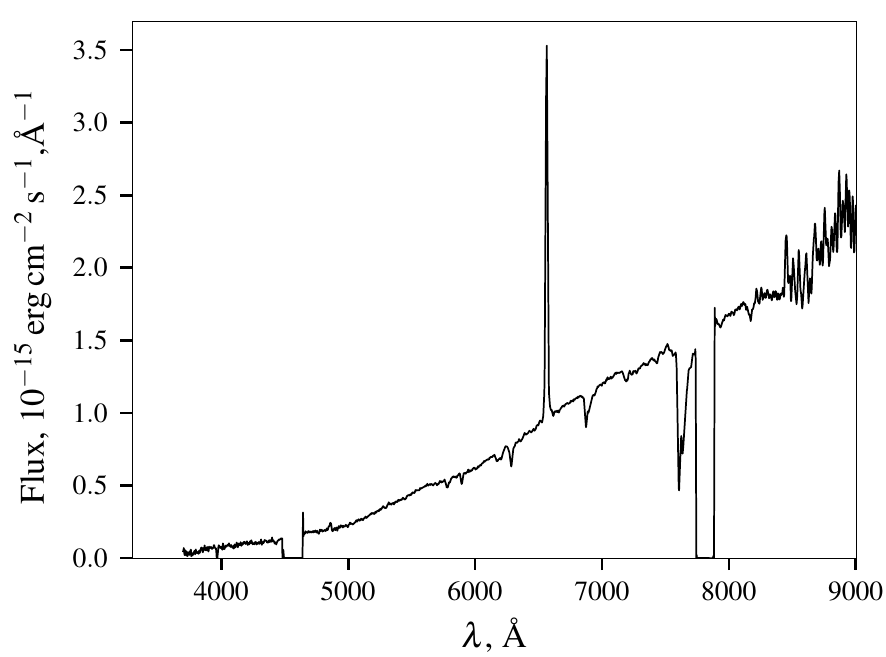}
    \caption{\textit{Top panel:} VVV ($JHK_S$) image of the region around the X-ray source. The green circles indicate the source position as observed by \ero in eRASS1-3, the red contours represent the $1\sigma$ and $3\sigma$ uncertainties for the refined position obtained as described in the text. The 2MASS position of the suggested near-infrared counterpart is indicated by the red cross. 
    \textit{Bottom panel:} The spectrum of the optical counterpart as observed by SALT, note the prominent  H$\alpha$ line.}
    \label{fig:saltspe}
\end{figure}

To determine the X-ray position of the source, we considered the rectified source positions from three \ero surveys as independent measurements and combined them following the procedure described in \citet{2020A&A...643A..62D}.
This allowed to fix the final X-ray position as
$\alpha_\mathrm{J2000.0}=12^\mathrm{h}44^\mathrm{m}03\fs57$, $\delta_\mathrm{J2000.0}=-63\degr22\arcmin28\farcs1$ 
with an uncertainty of about $1\farcs4$ at $1\sigma$ or ${\sim}$3\farcs7 at $3\sigma$ confidence level (not accounting for potential systematic uncertainties (up to 5$\arcsec$) of individual \ero measurements). As evident from Fig.~\ref{fig:saltspe}, the estimated X-ray position coincides with the near-infrared source 2MASS~J12440380$-$6322320, which was thus suggested as the tentative optical counterpart of the X-ray source \citep{2021ATel14357....1B}. Considering the variability reported by \citet{2021ATel14361....1B}, we observed the object with SALT \citep{2021ATel14364....1B} to finally confirm this identification.

{
In particular, the proposed optical counterpart was observed using the low resolution 300\,l\,mm$^{-1}$ transmission grating, typically used for spectral classification of transients. 
The RSS spectra were reduced using the \textsc{PySALT} primary data reduction pipeline \citep{2012ascl.soft07010C}, which corrects for bias, gain and detector mosaic amplifier cross-talk. We then used \textsc{iraf}\footnote{Image Reduction and Analysis Facility: iraf.noao.edu} for the remaining data reduction steps, namely wavelength calibration, background subtraction extraction of the one-dimensional spectra and relative flux calibration (absolute flux calibration is impossible with SALT, due to its design).}

The combined SALT spectrum presented in Fig.~\ref{fig:saltspe} shows a strong H$\alpha$ emission line, a weaker H$\beta$ emission line and a largely featureless red continuum, rising to longer wavelengths. It is thus consistent with a highly reddened early-type Be star located in the Galactic plane at ${\sim}$6\,kpc \citep{2021ATel14357....1B} and strongly suggests that the X-ray source is likely indeed a Be X-ray binary \citep{2021ATel14364....1B}. 

{The H$\alpha$ line is single peaked, but shows a slight asymmetry, with the peak flux on the red side of the line centre. The equivalent width (EW) and FWHM of the line are measured as -54.2$\pm$2.7\AA{} and 17.55$\pm$0.48\AA, respectively. The H$\beta$ line EW is approximately -5\AA, although this is hard to measure precisely due to the lower S/N at the blue end of the spectrum. The H$\alpha$ EW is large compared to most other BeXRBs. Based on the proposed orbital period of 138 days  \citep{2021ATel14361....1B} and the well-known correlation between orbital period and H$\alpha$ EW in BeXRBs \citep{2011Ap&SS.332....1R}, the expected EW would be around -35\AA, so our measurement suggests that the disc has grown to be larger than expected for this orbital period so the system indeed can be expected to be active in X-rays.} 

\subsection{\textit{NuSTAR} data analysis}

In this paper we mainly focus on the \textit{NuSTAR} observation of the source obtained as part of the approved guest observer program targeting X-ray binaries discovered by \srg. The observation was triggered following the SALT observation and was specifically optimized to mitigate potential stray light contamination from the bright nearby X-ray pulsar GX~301$-$2. In particular, it was scheduled to occur outside of the regular ``pre-periastron'' flares observed from this X-ray binary \citep{2010A&A...515A..10D}. 

\begin{figure}
    \centering
    \includegraphics[width=\columnwidth]{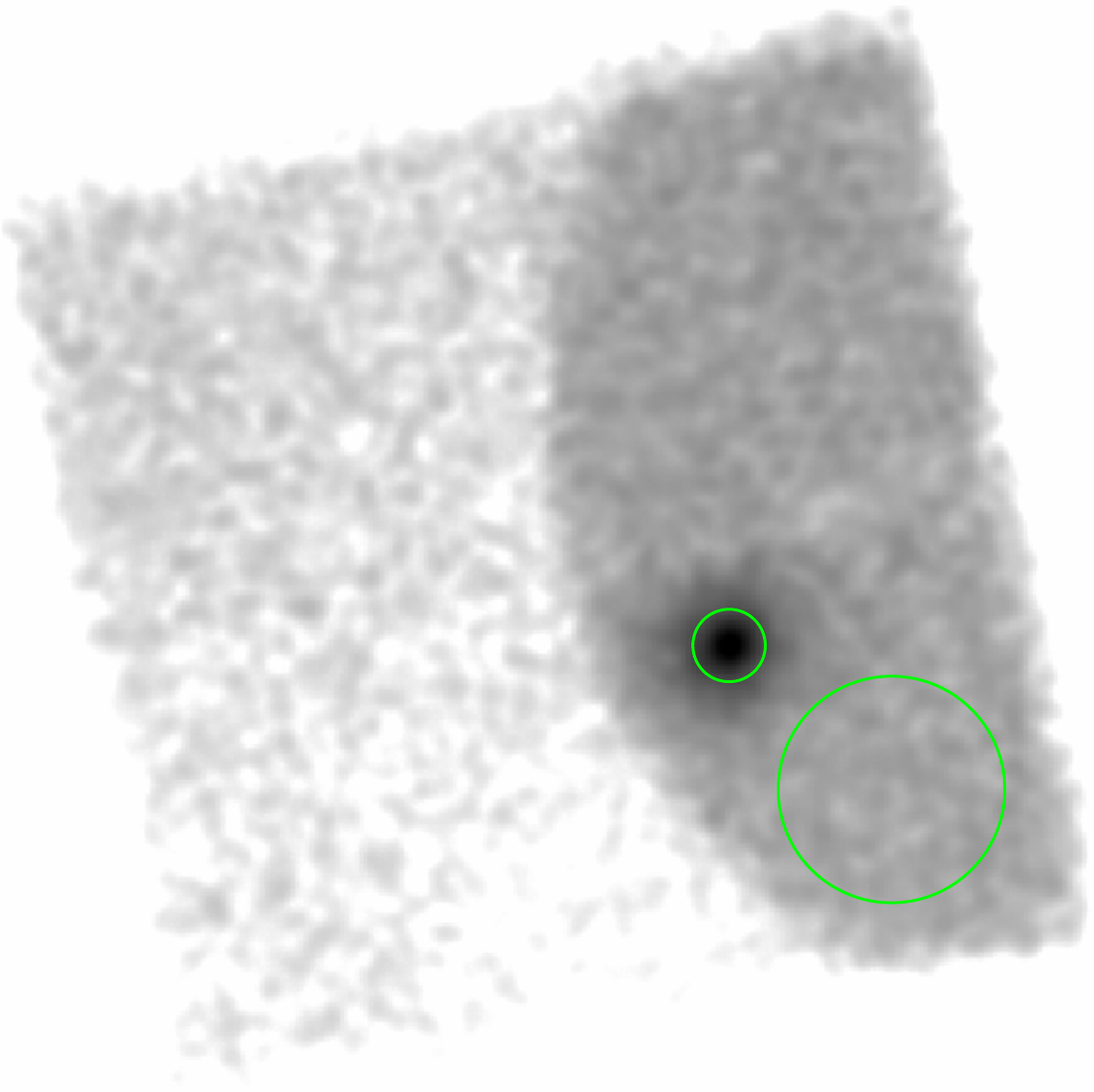}
    \caption{The smoothed plane image for the \textit{NuSTAR} FPMA detector unit in the 3-20\,keV energy band. Note the stray light contamination from GX~301$-$2. The locations of the source and background extraction regions are also shown (green circles).}
    \label{fig:stray}
\end{figure}

Unfortunately, as illustrated in Fig.~\ref{fig:stray}, it was not possible to avoid contamination completely. This resulted in a low signal to noise ratio for \srce, particularly in the hard band above 20\,keV. Therefore, for spectral and timing analysis in the hard band an optimized extraction region of 32\arcsec\ was chosen to achieve the optimal signal to noise ratio in the 20--80\,keV energy band as described by \cite{2018A&A...610A..88V}. However, taking into account that most of the photons are actually detected below 20\,keV, for timing analysis in the soft band a larger region (72\arcsec) was selected using the same procedure. Considering the apparent stray light contamination from GX 301$-$2, the background was also extracted from a contaminated area on the same chip away from the source using a circular region with radius of 100\arcsec. We verified that the choice of the location of the background region (as long as it is extracted from a stray light contaminated region) does not affect any of our conclusions.

\begin{figure}
    \centering
    \includegraphics[width=\columnwidth]{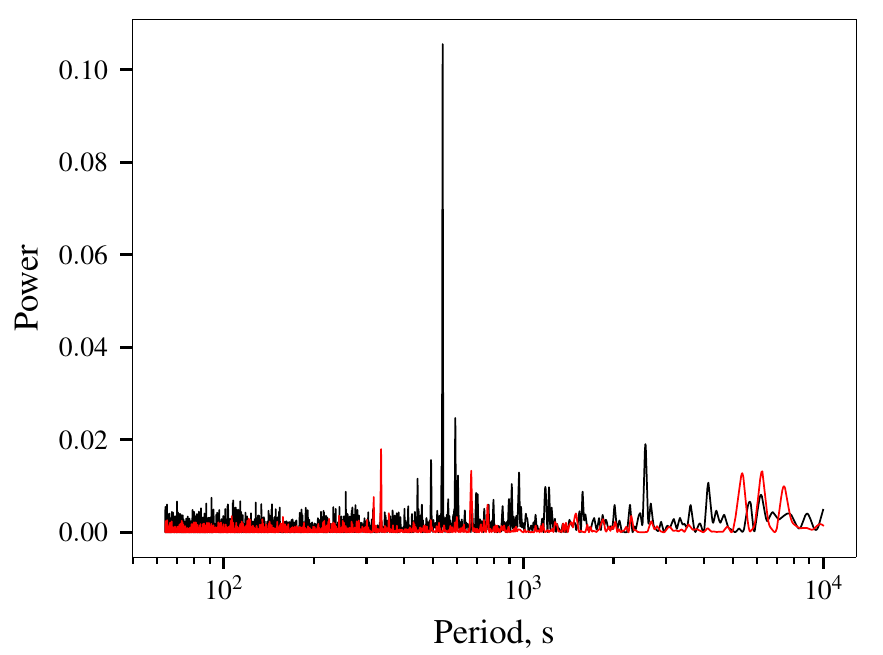}
    \caption{Lomb-Scargle periodogram for source (black) and background (red) light curves. Note that the $\sim$670\,s and $\sim$335\,s peaks from GX~301$-$2, which are visible in the background light curve are not present in the source light curve and do not overlap with the pulsed signal from \srca/\srce.}
    \label{fig:ls}
\end{figure}

\paragraph{Timing analysis}
To search for possible pulsations, we extracted a background-subtracted light curve in the 3--20\,keV energy band with a time resolution of 16\,s, combining the events extracted from both \textit{NuSTAR} units. We applied barycentric correction and then searched for pulsations using the generalized Lomb-Scargle periodogram \citep{2009A&A...496..577Z}. The result is presented in Fig.~\ref{fig:ls}, where a strong peak corresponding to a period of ${\sim}$538.56\,s is obvious (false detection probability as defined in \cite{2009A&A...496..577Z} is $\log{P}\sim-87$).

We note that this peak is not associated with the pulsations from the nearby GX~301$-$2, which has a longer period of ${\sim}$670\,s  that is clearly detected in the periodogram of the background light curve, but not for the background-subtracted light curve.
The detected pulsations imply that the X-ray source is an accreting neutron star, so we conclude \srca/\srce is likely indeed a new Be X-ray pulsar.

In order to refine the period value and in order to estimate its uncertainty, we used the phase-connection technique \citep{1981ApJ...247.1003D}. This approach allowed us to determine the final pulse period value as $P=538.70(5)$\,s. Note that this value does not account for Doppler delays associated with the orbital motion of the pulsar, as its orbital parameters are not yet known and can not be estimated using our comparatively short  duration of \textit{NuSTAR} observation (with respect to the expected orbital period).

\begin{figure}
    \centering
    \includegraphics[width=\columnwidth]{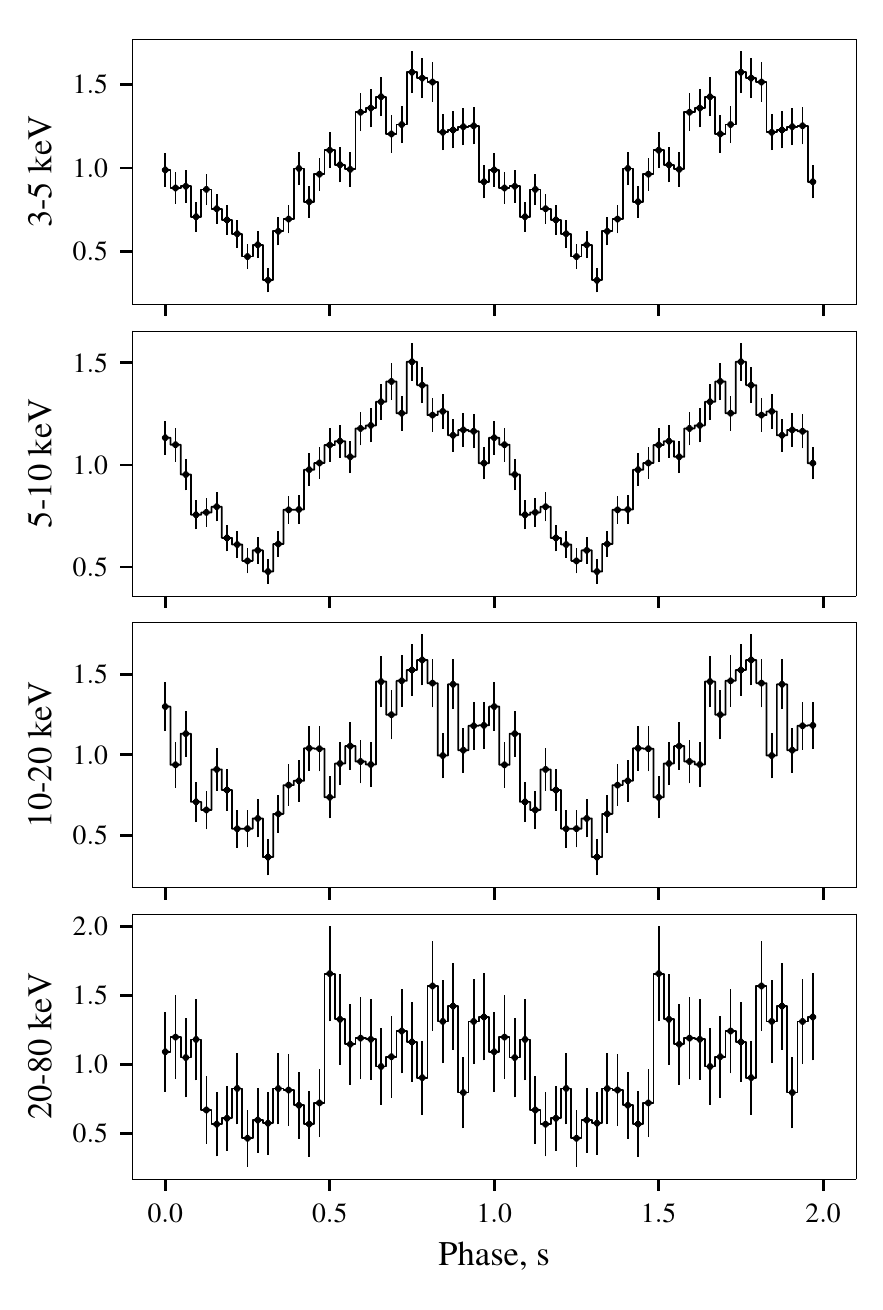}
    \caption{Normalized, background-subtracted pulse profiles of \srca/\srce as function of energy as observed by \textit{NuSTAR}.}
    \label{fig:pps}
\end{figure}

\begin{figure}
    \centering
    \includegraphics[width=\columnwidth]{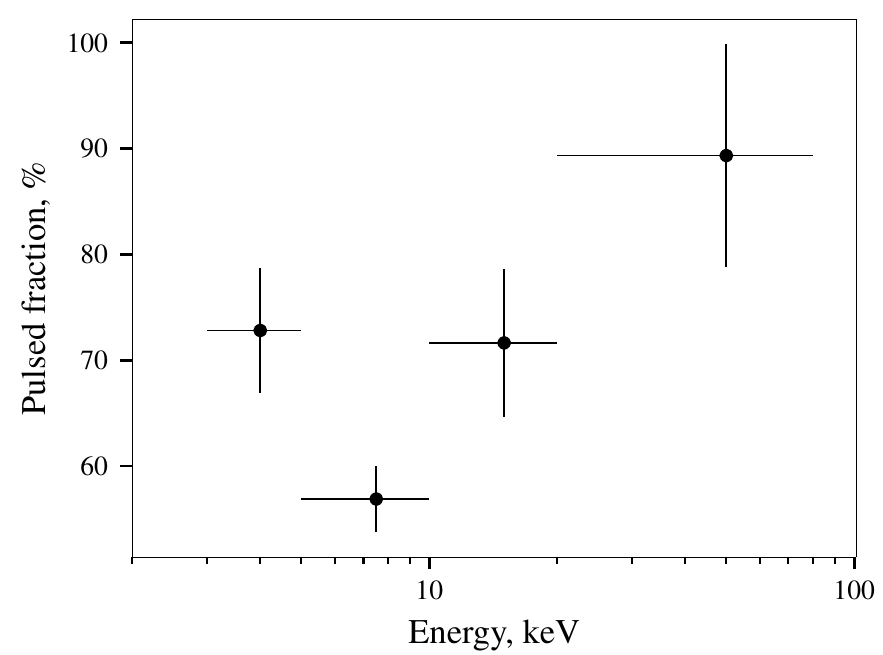}
    \caption{Pulsed fraction of \srca/\srce as function of energy as observed by \textit{NuSTAR}.}
    \label{fig:pf}
\end{figure}

To investigate the dependence of the pulse profiles of the source with energy, we extracted light curves in the 3--5, 5--10, 10--20, and 20--40\,keV energy bands, and folded those with the determined period using 32 equally spaced phase bins.
Again, the light curves from the two \textit{NuSTAR} units were co-added and background was subtracted. As is evident from Fig.~\ref{fig:pps}, the pulse profile is single peaked, sinusoidal, and exhibits no strong variations with energy. 
The pulsed fraction, defined here as $\max(r)-\min(r)/\max(r)+\min(r)$, where $r$ is the count rate, is high at all energies at ${\sim}$55--90\%  and, as illustrated in Fig.~\ref{fig:pf}, exhibits no clear increasing trend typical for X-ray pulsars \citep{2009AstL...35..433L} or other features apart from a slight drop in the 5--10\,keV band which is likely related to the chosen binning of the pulse profile and comparatively large measurement uncertainties.


\paragraph{Spectral analysis}
Taking into account that the stray light contamination is mostly relevant in the hard band, the broadband \textit{NuSTAR} spectrum was extracted using the region optimized for the hard band, i.e., using the extraction radius of 32\arcsec. The \srg spectra were extracted as described above. The \textit{Swift}/XRT spectrum of the source was extracted using the online tools provided by UKSSDC\footnote{\url{https://www.swift.ac.uk/user_objects/index.php}}\citep{2009MNRAS.397.1177E}. Considering the low statistics in all spectra with the exception of \textit{NuSTAR}, we modeled six data sets  (i.e., \ero spectra extracted using \texttt{srctool} task separately for the three surveys, the \art spectrum for the third survey,  the \textit{NuSTAR} spectra extracted separately from the two units, and the spectrum from the \textit{Swift}/XRT pointing carried out soon after \textit{NuSTAR}) using the same model including a cross-normalization constant to account for differences in flux. As already mentioned, this includes contemporary observations by \ero and \art which are, in fact, not strictly simultaneous due to the different field of view resulting in different length of scans over the source for the two instruments.
All spectra were grouped to contain at least one count per energy bin and modeled using Xspec v12.11.1 using the Cash \citep{1979ApJ...228..939C} and Anderson-Darling \citep{10.1214/aoms/1177729437} as fit and test statistics. This necessitated also the use of Monte-Carlo sampling for the final estimates of parameter uncertainties and the goodness of fit (using the \texttt{chain} and \texttt{goodness} commands in \texttt{Xspec}).

We considered several phenomenological models typically used to describe the spectra of X-ray pulsars to model the broadband spectrum of \srca/\srce. In particular, we first attempted to model the spectrum with absorbed  cutoff power-law models, i.e. \texttt{cutoffpl} and \texttt{highecut} models multiplied by a cross-normalization constant to account for differences in absolute flux calibration of individual instruments and flux variations between observations and \texttt{TBabs} \citep{2000ApJ...542..914W} to account for absorption. However, in both cases the cutoff energy was not constrained effectively reducing both models to a simple power law, which is unexpected for an X-ray pulsar. Nevertheless, the best-fit parameters for the absorbed power-law model are listed in Table~\ref{tab:spe}. We note that the goodness of fit assessed using a set of spectra simulated based on the best-fit power-law model using the \texttt{goodness} command turned out to be not acceptable (with over 95\% simulations yielding lower statistics than the data), so other models need be considered.

We note that the estimated broadband un-absorbed source flux for the power-law model of $\sim2.3\times10^{-11}$\,erg\,cm$^{-2}$\,s$^{-1}$ implies a luminosity of $\sim10^{35}$\,erg\,s$^{-1}$ for the assumed distance of 5.8\,kpc \citep{2021AJ....161..147B}, i.e., \srca/\srce appears to be more similar to the low-luminosity pulsars like X~Persei than to classical Be transients. We considered, therefore, also a model consisting of two comptonization components represented by the \texttt{CompTT} model \citep{1994ApJ...434..570T} with linked seed temperatures used to describe spectra of X~Persei and other similar sources \citep{2012A&A...540L...1D,2019MNRAS.483L.144T,2019MNRAS.487L..30T,2019A&A...621A.134T,2020arXiv200613596M,2021arXiv210110834D,2021arXiv210305728L}. The continuum was again modified to account for interstellar absorption using a \texttt{TBabs} component multiplied by a cross-normalization constant to account for flux variations and differences in absolute flux calibration. The best-fit results are presented in Table~\ref{tab:spe}. Note that the fit is not only statistically preferred to the power-law model, but more importantly, it is also statistically acceptable (verified using the \texttt{goodness} command). To compare the two models we used here the Bayesian information criterion  \citep[BIC, see i.e.][]{doi:10.1080/01621459.1995.10476572}, which indeed indicates a ``very strong'' preference for the Comptonization model (see Table~\ref{tab:spe} for BIC values). The best-fit parameters are also similar to those observed in other X-ray pulsars observed in a low-luminosity state (see, e.g., the papers cited above). We conclude, therefore, that \srca/\srce indeed shows a two-component spectrum similar to that of other X-ray pulsars observed at comparable luminosities as described in detail in Section~3.

Finally, comparison of spectra normalization obtained by various instruments also allows to asses variability of the source on long timescales. As illustrated in Table~\ref{tab:spe}, the source flux appears to be variable by a factor of six between the observations. We emphasize, however, that due to the short duration of the individual scans compared to spin period of the source, and high pulsed fraction revealed by \textit{NuSTAR}, most of this variability is likely associated with the short-term variability over the pulse period rather than long-term flux changes. We conclude, therefore, that long-term flux of the source remains comparatively stable with no clear outbursts typical for BeXRBs detected up to now.

\begin{table}[]
    \caption{Broadband spectral modeling results for \srca/\srce using absorbed power law (\texttt{powerlaw}) and two-component comptonization models (\texttt{CompTT2}).}
    \label{tab:spe}
    \centering
    \begin{tabular}{lll}
        \hline\noalign{\smallskip}
          & \texttt{powerlaw} & \texttt{CompTT2}  \\
        \hline\noalign{\smallskip}
         $C_{FMPB}$ & 1.08(3) & 1.08(4) \\
         $C_{XRT}$ & 0.6(1) & 0.6(1) \\
         $C_{e1}$ & 0.6(1) & 0.6(1) \\
         $C_{e2}$ & 0.16(9) &  0.17(4) \\
         $C_{e3}$ & 1.3(4)  & 1.3(3)\\
         $C_{ART}$ & 3.3(7)  & 2.9(3)\\
         $N_H$, $10^{22}$\,cm$^{-2}$ & 4.1(7) & 2.6(3)\\
         $\Gamma$, keV & 2.11(4) & \\
         $kT_0$, keV & & 0.12(3)\\
         $kT_1/kT_2$, keV & & 1.85(8)/8.3(8) \\
         $\tau_1/\tau_2$ & & 20(1)/13(1) \\
         $A_{\Gamma/1/2}/10^{-3}$ &  2.6(3) & 1.6(2)/0.08(1)\\
         $F_{\rm obs/src}$, 10$^{-11}$\,erg\,cm$^{-2}$\,s$^{-1}$ & 1.14/2.26 & 1.0/1.3 \\
         \hline\noalign{\smallskip}
         $\log{P_{AD}}$/dof & -8.81/1432 & -9.89/1427\\
         BIC & 46 & 64 \\
         \hline
    \end{tabular}
    \tablefoot{Test statistics probability ($\log{P_{AD}}$) and Bayesian information criterion (BIC) values are quoted.}
\end{table}

\begin{figure}
    \centering
    \includegraphics[width=\columnwidth]{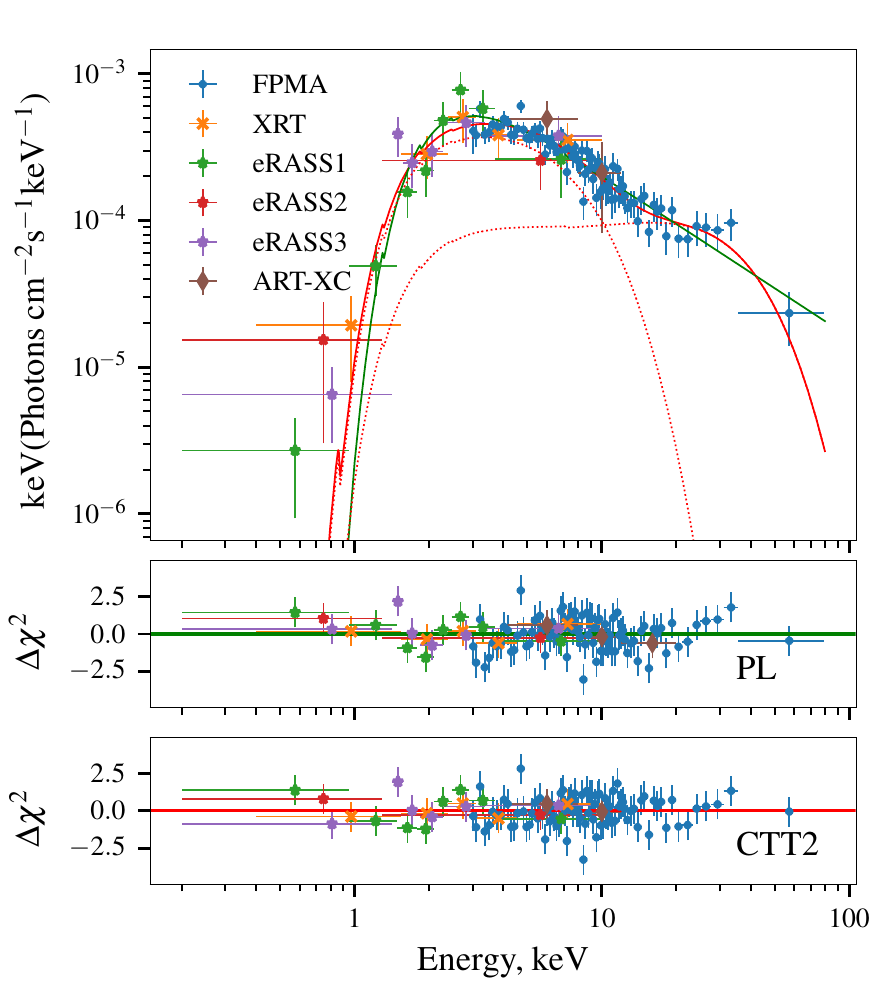}
    \caption{Unfolded spectrum of the source assuming the best-fit \texttt{CompTT2} model. For plotting all spectra were re-binned (to 10 (\ero and \textit{Swift}/XRT) or 25 counts (\textit{NuSTAR}) per energy bin) and scaled using the best-fit cross-normalization constants listed in Table~\ref{tab:spe}. Also for clarity only \textit{FPMA} data for \textit{NuSTAR} is shown as spectra from both modules look nearly identical.}
    \label{fig:spe}
\end{figure}

\section{Discussion}

\begin{figure}
    \centering
    \includegraphics[width=\columnwidth]{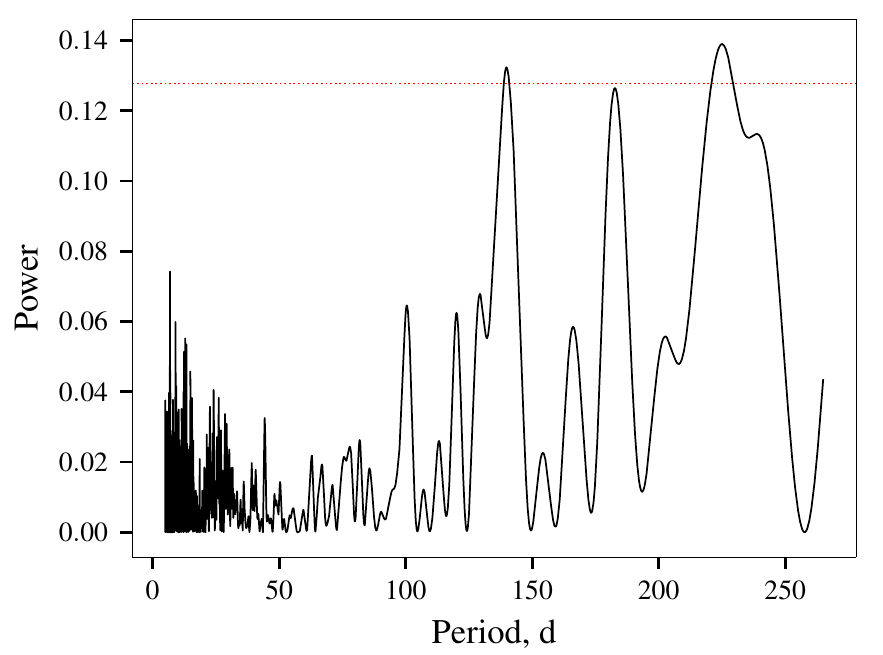}
    \includegraphics[width=\columnwidth]{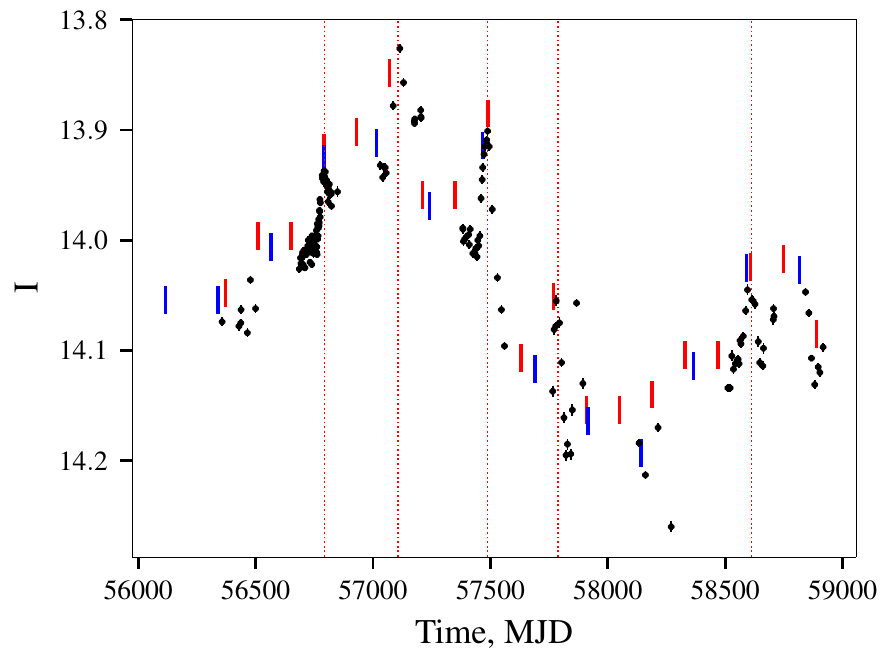}
    \caption{Top panel: generalized Lomb-Scargle periodogram for OGLE $I$-band ligthcurve of \srca/\srce counterpart. The red line indicates the expected $3\sigma$ upper limit on noise amplitude. Two peaks exceeding the noise level correspond to periods of $\sim$138 and $\sim$225\,d. Bottom panel: OGLE $I$-band light curve (black points). Note the five short, quasi-periodic flares indicated with red dotted lines. The short vertical lines indicate expected flare timings assuming that they are related to orbital phase, with either 138\,d (red) or 225\,d (blue) orbital period.}
    \label{fig:ogle}
\end{figure}
\begin{figure}
    \centering
    \includegraphics[width=\columnwidth]{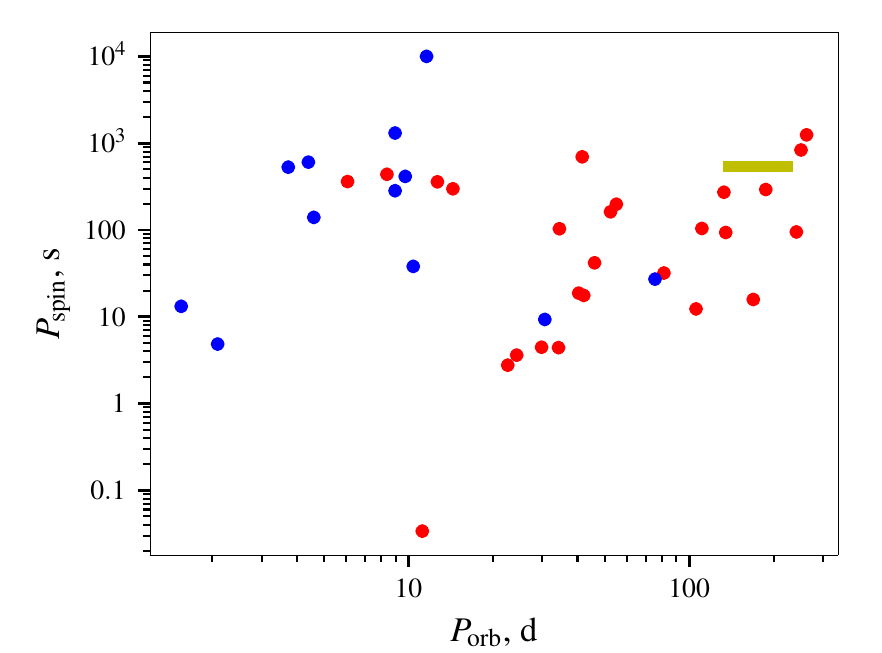}
    \caption{Orbital period - spin period \citep{1986MNRAS.220.1047C} diagram for wind accreting (blue) and Be (red) high-mass X-ray binaries. The position of \srca/\srce is indicated with the yellow line assuming the orbital period from 138 to 225\,d.}
    \label{fig:corbet}
\end{figure}

As discussed above, the observed properties of the newly discovered BeXRB pulsar \srce are rather similar to those of other long-period pulsars observed at low luminosities. In particular, similar spectra were reported from the transient XRBs 1A~0535+262 \citep{2019MNRAS.487L..30T} and GX~304$-$1 \citep{2019MNRAS.483L.144T} when observed in quiescence. Another object where a similar X-ray spectrum was observed is the only persistent BeXRB known, i.e. X~Persei, which historically was only observed in the low-luminosity state, presumably due to the longer orbital period and wider separation between the neutron star and the donor. 

At first sight \srce appears to be similar to that source, i.e. likely a low-luminosity persistent pulsar. The relatively long orbital period of $\sim$138\,d suggested by \cite{2021ATel14361....1B} and the non-detection of strong outbursts, despite the observed prominent H$\alpha$ emission line indicating the presence of an extended decretion disk around the primary \citep{2016A&A...590A.122R}, suggests that the pulsar probably is indeed far from the companion and thus unlikely to cross the circum-stellar disk to undergo strong outbursts. For long-period pulsars the accretion is expected to proceed from the ``cold'' non-ionized accretion disk regularly replenished by the  wind of the primary \citep{2017A&A...608A..17T}. As a result, a  relatively stable luminosity of $\sim10^{34-35}$\,erg\,cm$^{-2}$\,s$^{-1}$, consistent with the observations, might be expected to be maintained for a long time.

We note, however, that the orbital period of the source is not yet known for sure, so it can not be fully excluded that the object will eventually undergo an outburst. Indeed, the suggestion of the orbital period by \cite{2021ATel14361....1B} is based on the detection of  several short episodes when the companion brightened by ${\sim}$0.1\,mag (illustrated in Fig.~\ref{fig:ogle}).
However, considering that only five such episodes can be identified in the light curve, one can not be fully sure that those are periodic and associated with orbital motion.
A search for periodicity in the raw OGLE light curve reveals several marginally significant peaks, and the peak at $\sim$140\,d, i.e., slightly longer than the period suggested by \citet{2021ATel14361....1B}, is only one of them. In particular, the longer period of $\sim$225\,d also appears to be consistent with most of the brightening episodes (see i.e. Fig.~\ref{fig:ogle}).  Furthermore, the 225\,d period also better fits the correlation between H$\alpha$ EW and orbital period (discussed in Sect. 2.1). This period corresponds to the expected EW of $\sim$50\,\AA\  \citep{2011Ap&SS.332....1R}, consistent with our measurement with SALT.
Both suggested values  are also consistent with expectations based on the relation between spin and orbital periods for X-ray pulsars \citep{1986MNRAS.220.1047C} as illustrated in Fig.~\ref{fig:corbet} (here shown based on the data reported in the catalog of the Galactic HMXBs by \citealt{2006A&A...455.1165L}). Further monitoring of the source in X-ray and optical bands is, therefore, required to clarify the long-term behaviour of the source.

\section{Conclusions}
\srca/\srce is one of the first new Galactic X-ray pulsars identified in the all-sky X-ray survey of \srg (Lutovinov et al., submitted), and the first confirmed Galactic Be X-ray binary (note that several such objects also have been already detected  in the Large Magellanic Cloud where longer exposures are available, see e.g. \citealt{2021A&A...647A...8M,2021arXiv210303657M,2020ATel13609....1H,2020ATel13610....1M}). Although it is not yet fully clear whether the source is a persistent low-luminous pulsar similar to X~Persei or a low duty cycle transient Be X-ray binary. The non-detection of a strong outburst with peak luminosity of $\sim$10$^{36-37}$\,erg\,s$^{-1}$  typical for other BeXRBs despite the observed prominent H$\alpha$ emission line from the companion, which indicates the presence of an expanded decretion disk \citep{2016A&A...590A.122R} around the donor suggests, however, that the former is more likely. We note that pre-launch estimates indicate that several tens of objects like this one are expected to be detected in the \ero survey \citep{2014A&A...567A...7D}, and \srca/\srce might be the first of them. In fact, only \srg can unveil the population of such faint persistent objects and we are searching for more of those in the survey. 

\begin{acknowledgements}
This work is based on data from \ero, the primary instrument aboard SRG, a joint Russian-German science mission supported by the Russian Space Agency (Roskosmos), in the interests of the Russian Academy of Sciences represented by its Space Research Institute (IKI), and the Deutsches Zentrum für Luft- und Raumfahrt (DLR). The SRG spacecraft was built by Lavochkin Association (NPOL) and its subcontractors, and is operated by NPOL with support from the Space Research Institute (IKI) and Max Planck Institute for Extraterrestrial Physics (MPE).

The development and construction of the \ero X-ray instrument was led by MPE, with contributions from the Dr. Karl Remeis Observatory Bamberg \& ECAP (FAU Erlangen-Nuernberg), the University of Hamburg Observatory, the Leibniz Institute for Astrophysics Potsdam (AIP), and the Institute for Astronomy and Astrophysics of the University of Tübingen, with the support of DLR and the Max Planck Society. The Argelander Institute for Astronomy of the University of Bonn and the Ludwig Maximilians Universität Munich also participated in the science preparation for \ero.

The \ero data shown here were processed using the eSASS/NRTA software system developed by the German \ero consortium.

The {\it Mikhail Pavlinsky} \art\ telescope is  the  hard X-ray instrument on board the \srg\ observatory, a flagship astrophysical project of the Russian Federal Space Program realized by the Russian Space Agency, in the interests of the Russian Academy of Sciences. \art\ was developed by the Space Research Institute (IKI, Moscow) and the Russian Federal Nuclear Center -- All-Russian Scientific Research Institute for Experimental Physics (RFNC-VNIIEF, Sarov) with the participation of the  NASA's Marshall Space Flight Center (MSFC). The \art\ team thanks the Russian Space Agency, Russian Academy of Sciences and State Corporation Rosatom for the support of the \srg\ project and \art\ telescope. 

We thank the \textit{NuSTAR} SOC for their invaluable help in the quick scheduling of the observations. This research has made use of data, software and/or web tools obtained from the High Energy Astrophysics Science Archive Research Center (HEASARC), a service of the Astrophysics Science Division at NASA/GSFC and of the Smithsonian Astrophysical Observatory's High Energy Astrophysics Division. 

This work made use of data supplied by the UK Swift Science Data Centre at the University of Leicester. AS, IM and AL acknowledge support from Russian Science Foundation via grant 19-12-00423. The SALT observations reported here were triggered by SRG and obtained through the SALT Large Science program (2018-2-LSP-001; PI: Buckley). Polish participation in SALT is funded by grant no. MNiSW DIR/WK/2016/07. DAHB is supported by the National Research Foundation (NRF) of South Africa. MG is supported by the EU Horizon 2020 research and innovation programme under grant agreement No 101004719.
 
\end{acknowledgements}

\vspace{-0.3cm}
\bibliography{biblio}   

\begin{thebibliography}{44}
\expandafter\ifx\csname natexlab\endcsname\relax\def\natexlab#1{#1}\fi

\bibitem[{Anderson \& Darling(1952)}]{10.1214/aoms/1177729437}
Anderson, T.~W. \& Darling, D.~A. 1952, The Annals of Mathematical Statistics,
  23, 193

\bibitem[{{Bailer-Jones} {et~al.}(2021){Bailer-Jones}, {Rybizki}, {Fouesneau},
  {Demleitner}, \& {Andrae}}]{2021AJ....161..147B}
{Bailer-Jones}, C.~A.~L., {Rybizki}, J., {Fouesneau}, M., {Demleitner}, M., \&
  {Andrae}, R. 2021, \aj, 161, 147

\bibitem[{{Boller} {et~al.}(2016){Boller}, {Freyberg}, {Tr{\"u}mper}, {Haberl},
  {Voges}, \& {Nandra}}]{2016A&A...588A.103B}
{Boller}, T., {Freyberg}, M.~J., {Tr{\"u}mper}, J., {et~al.} 2016, \aap, 588,
  A103

\bibitem[{{Buckley}(2015)}]{Buckley2015}
{Buckley}, D.~A.~H. 2015, in Proceeding of Science, SALT Science Conference
  2015 (SSC2015; https://pos.sissa.it/250/021/pdf), 1--14

\bibitem[{{Buckley} {et~al.}(2021){Buckley}, {Gromadzki}, {Townsend},
  {Monageng}, {Lutovinov}, {Doroshenko}, \& {Rau}}]{2021ATel14364....1B}
{Buckley}, D.~A.~H., {Gromadzki}, M., {Townsend}, L., {et~al.} 2021, The
  Astronomer's Telegram, 14364, 1

\bibitem[{{Buckley} {et~al.}(2006){Buckley}, {Swart}, \&
  {Meiring}}]{2006SPIE.6267E..0ZB}
{Buckley}, D.~A.~H., {Swart}, G.~P., \& {Meiring}, J.~G. 2006, in \procspie,
  Vol. 6267, Society of Photo-Optical Instrumentation Engineers (SPIE)
  Conference Series, 62670Z

\bibitem[{{Cash}(1979)}]{1979ApJ...228..939C}
{Cash}, W. 1979, \apj, 228, 939

\bibitem[{{Corbet}(1986)}]{1986MNRAS.220.1047C}
{Corbet}, R.~H.~D. 1986, \mnras, 220, 1047

\bibitem[{{Crawford} {et~al.}(2012){Crawford}, {Still}, {Schellart}, {Balona},
  {Buckley}, {Gulbis}, {Kniazev}, {Kotze}, {Loaring}, {Nordsieck}, {Pickering},
  {Potter}, {Romero Colmenero}, {Vaisanen}, {Wiliams}, \&
  {Zietsman}}]{2012ascl.soft07010C}
{Crawford}, S.~M., {Still}, M., {Schellart}, P., {et~al.} 2012, {PySALT: SALT
  science pipeline}, Astrophysics Source Code Library

\bibitem[{{Deeter} {et~al.}(1981){Deeter}, {Boynton}, \&
  {Pravdo}}]{1981ApJ...247.1003D}
{Deeter}, J.~E., {Boynton}, P.~E., \& {Pravdo}, S.~H. 1981, \apj, 247, 1003

\bibitem[{{Dennerl} {et~al.}(2020){Dennerl}, {Andritschke}, {Br{\"a}uninger},
  {Burkert}, {Burwitz}, {Emberger}, {Freyberg}, {Friedrich}, {Gaida},
  {Granato}, \& et~al.}]{2020SPIE11444E..4QD}
{Dennerl}, K., {Andritschke}, R., {Br{\"a}uninger}, H., {et~al.} 2020, in
  Society of Photo-Optical Instrumentation Engineers (SPIE) Conference Series,
  Vol. 11444, Society of Photo-Optical Instrumentation Engineers (SPIE)
  Conference Series, 114444Q

\bibitem[{{Doroshenko} {et~al.}(2014){Doroshenko}, {Ducci}, {Santangelo}, \&
  {Sasaki}}]{2014A&A...567A...7D}
{Doroshenko}, V., {Ducci}, L., {Santangelo}, A., \& {Sasaki}, M. 2014, \aap,
  567, A7

\bibitem[{{Doroshenko} {et~al.}(2012){Doroshenko}, {Santangelo}, {Kreykenbohm},
  \& {Doroshenko}}]{2012A&A...540L...1D}
{Doroshenko}, V., {Santangelo}, A., {Kreykenbohm}, I., \& {Doroshenko}, R.
  2012, \aap, 540, L1

\bibitem[{{Doroshenko} {et~al.}(2010){Doroshenko}, {Santangelo}, {Suleimanov},
  {Kreykenbohm}, {Staubert}, {Ferrigno}, \& {Klochkov}}]{2010A&A...515A..10D}
{Doroshenko}, V., {Santangelo}, A., {Suleimanov}, V., {et~al.} 2010, \aap, 515,
  A10

\bibitem[{{Doroshenko} {et~al.}(2021){Doroshenko}, {Santangelo}, {Tsygankov},
  \& {Ji}}]{2021arXiv210110834D}
{Doroshenko}, V., {Santangelo}, A., {Tsygankov}, S., \& {Ji}, L. 2021, arXiv
  e-prints, arXiv:2101.10834

\bibitem[{{Doroshenko} {et~al.}(2020){Doroshenko}, {Suleimanov}, {Tsygankov},
  {M{\"o}nkk{\"o}nen}, {Ji}, \& {Santangelo}}]{2020A&A...643A..62D}
{Doroshenko}, V., {Suleimanov}, V., {Tsygankov}, S., {et~al.} 2020, \aap, 643,
  A62

\bibitem[{{Evans} {et~al.}(2009){Evans}, {Beardmore}, {Page}, {Osborne},
  {O'Brien}, {Willingale}, {Starling}, {Burrows}, {Godet}, {Vetere}, \&
  et~al.}]{2009MNRAS.397.1177E}
{Evans}, P.~A., {Beardmore}, A.~P., {Page}, K.~L., {et~al.} 2009, \mnras, 397,
  1177

\bibitem[{{Freyberg} {et~al.}(2020){Freyberg}, {Perinati}, {Pacaud}, {Eraerds},
  {Churazov}, {Dennerl}, {Predehl}, {Merloni}, {Meidinger}, {Bulbul}, \&
  et~al.}]{2020SPIE11444E..1OF}
{Freyberg}, M., {Perinati}, E., {Pacaud}, F., {et~al.} 2020, in Society of
  Photo-Optical Instrumentation Engineers (SPIE) Conference Series, Vol. 11444,
  Society of Photo-Optical Instrumentation Engineers (SPIE) Conference Series,
  114441O

\bibitem[{{Haberl} {et~al.}(2020){Haberl}, {Maitra}, {Carpano}, {Ducci},
  {Doroshenko}, {Koenig}, {Buckley}, {Monageng}, \&
  {Udalski}}]{2020ATel13609....1H}
{Haberl}, F., {Maitra}, C., {Carpano}, S., {et~al.} 2020, The Astronomer's
  Telegram, 13609, 1

\bibitem[{{Harrison} {et~al.}(2013){Harrison}, {Craig}, {Christensen},
  {Hailey}, {Zhang}, {Boggs}, {Stern}, {Cook}, {Forster}, {Giommi}, \&
  et~al.}]{2013ApJ...770..103H}
{Harrison}, F.~A., {Craig}, W.~W., {Christensen}, F.~E., {et~al.} 2013, \apj,
  770, 103

\bibitem[{Kass \& Raftery(1995)}]{doi:10.1080/01621459.1995.10476572}
Kass, R.~E. \& Raftery, A.~E. 1995, Journal of the American Statistical
  Association, 90, 773

\bibitem[{{Liu} {et~al.}(2006){Liu}, {van Paradijs}, \& {van den
  Heuvel}}]{2006A&A...455.1165L}
{Liu}, Q.~Z., {van Paradijs}, J., \& {van den Heuvel}, E.~P.~J. 2006, \aap,
  455, 1165

\bibitem[{{Lutovinov} {et~al.}(2021){Lutovinov}, {Tsygankov}, {Molkov},
  {Doroshenko}, {Mushtukov}, {Arefiev}, {Lapshov}, {Tkachenko}, \&
  {Pavlinsky}}]{2021arXiv210305728L}
{Lutovinov}, A., {Tsygankov}, S., {Molkov}, S., {et~al.} 2021, arXiv e-prints,
  arXiv:2103.05728

\bibitem[{{Lutovinov} {et~al.}(2013){Lutovinov}, {Revnivtsev}, {Tsygankov}, \&
  {Krivonos}}]{2013MNRAS.431..327L}
{Lutovinov}, A.~A., {Revnivtsev}, M.~G., {Tsygankov}, S.~S., \& {Krivonos},
  R.~A. 2013, \mnras, 431, 327

\bibitem[{{Lutovinov} \& {Tsygankov}(2009)}]{2009AstL...35..433L}
{Lutovinov}, A.~A. \& {Tsygankov}, S.~S. 2009, Astronomy Letters, 35, 433

\bibitem[{{Maitra} {et~al.}(2020){Maitra}, {Haberl}, {Carpano}, {Koenig},
  {Doroshenko}, {Ducci}, {Buckley}, {Monageng}, \&
  {Udalski}}]{2020ATel13610....1M}
{Maitra}, C., {Haberl}, F., {Carpano}, S., {et~al.} 2020, The Astronomer's
  Telegram, 13610, 1

\bibitem[{{Maitra} {et~al.}(2021{\natexlab{a}}){Maitra}, {Haberl}, {Maggi},
  {Kavanagh}, {Vasilopoulos}, {Sasaki}, {Filipovic}, \&
  {Udalski}}]{2021arXiv210303657M}
{Maitra}, C., {Haberl}, F., {Maggi}, P., {et~al.} 2021{\natexlab{a}}, arXiv
  e-prints, arXiv:2103.03657

\bibitem[{{Maitra} {et~al.}(2021{\natexlab{b}}){Maitra}, {Haberl},
  {Vasilopoulos}, {Ducci}, {Dennerl}, \& {Carpano}}]{2021A&A...647A...8M}
{Maitra}, C., {Haberl}, F., {Vasilopoulos}, G., {et~al.} 2021{\natexlab{b}},
  \aap, 647, A8

\bibitem[{{Merloni} {et~al.}(2012){Merloni}, {Predehl}, {Becker},
  {B{\"o}hringer}, {Boller}, {Brunner}, {Brusa}, {Dennerl}, {Freyberg},
  {Friedrich}, \& et~al.}]{2012arXiv1209.3114M}
{Merloni}, A., {Predehl}, P., {Becker}, W., {et~al.} 2012, arXiv e-prints,
  arXiv:1209.3114

\bibitem[{Mroz \& Udalski(2021)}]{2021ATel14361....1B}
Mroz, P. \& Udalski, A. 2021, The Astronomer's Telegram, 14361, 1

\bibitem[{{Mushtukov} {et~al.}(2020){Mushtukov}, {Suleimanov}, {Tsygankov}, \&
  {Portegies Zwart}}]{2020arXiv200613596M}
{Mushtukov}, A.~A., {Suleimanov}, V.~F., {Tsygankov}, S.~S., \& {Portegies
  Zwart}, S. 2020, arXiv e-prints, arXiv:2006.13596

\bibitem[{{Pavlinsky} {et~al.}(2021){Pavlinsky}, {Tkachenko}, {Levin},
  {Alexandrovich}, {Arefiev}, {Babyshkin}, {Batanov}, {Bodnar}, {Bogomolov},
  {Bubnov}, \& et~al.}]{2021arXiv210312479P}
{Pavlinsky}, M., {Tkachenko}, A., {Levin}, V., {et~al.} 2021, arXiv e-prints,
  arXiv:2103.12479

\bibitem[{{Predehl} {et~al.}(2021){Predehl}, {Andritschke}, {Arefiev},
  {Babyshkin}, {Batanov}, {Becker}, {B{\"o}hringer}, {Bogomolov}, {Boller},
  {Borm}, \& et~al.}]{2021A&A...647A...1P}
{Predehl}, P., {Andritschke}, R., {Arefiev}, V., {et~al.} 2021, \aap, 647, A1

\bibitem[{{Reig}(2011)}]{2011Ap&SS.332....1R}
{Reig}, P. 2011, \apss, 332, 1

\bibitem[{{Reig} {et~al.}(2016){Reig}, {Nersesian}, {Zezas}, {Gkouvelis}, \&
  {Coe}}]{2016A&A...590A.122R}
{Reig}, P., {Nersesian}, A., {Zezas}, A., {Gkouvelis}, L., \& {Coe}, M.~J.
  2016, \aap, 590, A122

\bibitem[{{Semena} {et~al.}(2021){Semena}, {Doroshenko}, {Arefiev},
  {Lutovinov}, Maitra, Mereminskiy, Molkov, Rau, Weber, \&
  Wilms}]{2021ATel14357....1B}
{Semena}, A., {Doroshenko}, V., {Arefiev}, A., {et~al.} 2021, The Astronomer's
  Telegram, 14357, 1

\bibitem[{{Titarchuk}(1994)}]{1994ApJ...434..570T}
{Titarchuk}, L. 1994, \apj, 434, 570

\bibitem[{{Tsygankov} {et~al.}(2019{\natexlab{a}}){Tsygankov}, {Doroshenko},
  {Mushtukov}, {Lutovinov}, \& {Poutanen}}]{2019A&A...621A.134T}
{Tsygankov}, S.~S., {Doroshenko}, V., {Mushtukov}, A.~A., {Lutovinov}, A.~A.,
  \& {Poutanen}, J. 2019{\natexlab{a}}, \aap, 621, A134

\bibitem[{{Tsygankov} {et~al.}(2019{\natexlab{b}}){Tsygankov}, {Doroshenko},
  {Mushtukov}, {Suleimanov}, {Lutovinov}, \& {Poutanen}}]{2019MNRAS.487L..30T}
{Tsygankov}, S.~S., {Doroshenko}, V., {Mushtukov}, A.~A., {et~al.}
  2019{\natexlab{b}}, \mnras, 487, L30

\bibitem[{{Tsygankov} {et~al.}(2017){Tsygankov}, {Mushtukov}, {Suleimanov},
  {Doroshenko}, {Abolmasov}, {Lutovinov}, \& {Poutanen}}]{2017A&A...608A..17T}
{Tsygankov}, S.~S., {Mushtukov}, A.~A., {Suleimanov}, V.~F., {et~al.} 2017,
  \aap, 608, A17

\bibitem[{{Tsygankov} {et~al.}(2019{\natexlab{c}}){Tsygankov}, {Rouco
  Escorial}, {Suleimanov}, {Mushtukov}, {Doroshenko}, {Lutovinov}, {Wijnands},
  \& {Poutanen}}]{2019MNRAS.483L.144T}
{Tsygankov}, S.~S., {Rouco Escorial}, A., {Suleimanov}, V.~F., {et~al.}
  2019{\natexlab{c}}, \mnras, 483, L144

\bibitem[{{Vybornov} {et~al.}(2018){Vybornov}, {Doroshenko}, {Staubert}, \&
  {Santangelo}}]{2018A&A...610A..88V}
{Vybornov}, V., {Doroshenko}, V., {Staubert}, R., \& {Santangelo}, A. 2018,
  \aap, 610, A88

\bibitem[{{Wilms} {et~al.}(2000){Wilms}, {Allen}, \&
  {McCray}}]{2000ApJ...542..914W}
{Wilms}, J., {Allen}, A., \& {McCray}, R. 2000, \apj, 542, 914

\bibitem[{{Zechmeister} \& {K{\"u}rster}(2009)}]{2009A&A...496..577Z}
{Zechmeister}, M. \& {K{\"u}rster}, M. 2009, \aap, 496, 577

\end{thebibliography}
\vspace{-0.3cm}

\end{document}